\documentstyle[revtex]{aps}
\begin{document}
\begin{title}
Stability and effective masses of
composite-fermions in the first and second Landau Level
\end{title}

\author{R. Morf$^1$ and N. d'Ambrumenil$^2$}
\begin{instit}
$^1$Paul Scherrer Institut, Badenerstrasse 569, CH 8048 Z\"urich, Switzerland
\end{instit}
\begin{instit}
$^2$Department of Physics, University of Warwick, Coventry, CV4 7AL, UK
\end{instit}

\begin{abstract}

We propose a measure of the stability of  composite fermions (CF's) at
even-denominator Landau-level filling fractions. Assuming
Landau-level mixing effects are not strong, we show
that the CF liquid at $\nu=2+1/2$ in the $n=1$ Landau
level cannot exist and relate this to the absence of a
hierarchy of incompressible states for filling fractions
$2+1/3 < \nu <  2+2/3$.
We find that a polarized CF liquid should exist at $\nu=2+1/4$.
We also show that, for CF states, the
variation with system size of the ground state
energy of interacting electrons  follows that for
non-interacting particles in zero magnetic field. We use this to
estimate the CF effective masses.

\end{abstract}
\pacs{PACS numbers:73.40H,03.70+k}

\narrowtext

Two-dimensional systems of electrons in magnetic fields, corresponding
to Landau-level filling fractions with even denominator, were for a
long time mysterious. Halperin, Lee and Read (HLR) partially cleared up this
mystery \cite{HLR},
when they identified the low-temperature phase of fully spin-polarised
electrons at filling fraction $\nu = 1/2$, as Fermi liquids of
composite fermions (CF's) \cite{Jain89,Lopez91}.
This theory accounted for
the available experimental observations \cite{WillettSAW,Willett93}
for systems at or close to $\nu =1/2$.
HLR pointed out that their CF formalism easily generalizes
to other filling fractions and to
higher Landau levels (LL's), where, if applicable, it
describes a Fermi liquid
phase for spin-polarised electrons at all even-denominator filling fractions.

Here we analyse microscopically the stability of the CF state. We propose
a definition of the `binding energy' of flux quanta to electrons,
which we argue puts an upper bound on the stability of
CF's. We also show that, although the formalism for CF generalizes to
higher LL's, the stability of the Fermi liquid state depends
strongly on LL index. This leads us to predict that
the polarised CF-liquid cannot be stable at $\nu = 5/2$.
We also predict that
at $\nu = 9/4$, the polarised CF-liquid will be stable and should show
up clearly in experiment. Our calculations suggest that this state is
more strongly stabilised than its counterpart in the lowest LL
at $\nu = 1/4$ (although it may not be much easier to observe,
because only 1/9 of the electrons would actually form the correlated
state.) Finally, we use the variation with system size of the
ground state energy to estimate the CF effective
masses.

Our results leave open the question of what is
the nature of the polarized state at $\nu=5/2$.
It has been known for some time that the behaviour at filling
fractions $\nu = 5/2$ is anomalous, with some samples showing
a plateau in the Hall conductivity, $\sigma_{xy}$ \cite{Willett5/2,Morf5/2}.
These anomalies disappear in a tilted magnetic field and
are thought to be associated with an incompressible non-fully polarized
ground state \cite{Eisenstein88,HaldRez88}, which is destabilized by a
sufficiently large
Zeeman energy. Our results show that the polarized ground state at $\nu=5/2$,
which forms in this large Zeeman/high density limit,
will not be a simple CF-liquid, at least in the absence of strong
LL-mixing.
(Although preliminary surface acoustic wave measurements \cite{Willett93}
show no evidence at $\nu=5/2$ for the `free fermion-like' resonances
expected for CF's, this may reflect a non-fully polarized
incompressible ground state.)

A trial wavefunction for the CF liquid ground state at $\nu = 1/2$
has been written down for $N$ electrons in the lowest ($n=0$) LL
\cite{RR94}. In
the symmetric gauge ${\bf A} = (B/2)(y,-x,0)$ and using the complex
position coordinate for the $j$'th electron $z_j = x_j + iy_j$, it is
\begin{eqnarray}
\psi_{CF}^{n=0}
& = & P_{n=0} \mbox{det} |e^{i{\bf k}_{\alpha}{\bf r}_i}| \psi_2^{(0)}
\label{eq:psiCF} \\
\psi_2^{(0)} & = & \prod_{i<j}^N (z_i - z_j)^2 \exp{-\sum_k |z_k|^2/4l_0^2}.
\label{eq:psi2}
\end{eqnarray}
The operator $P_{n=0}$ projects the wavefunction on the $n=0$ LL.
$l_0$ is the magnetic length.
The factor $\psi_2^{(0)}$ takes care of the singular gauge transformation
which describes the formation of the CF's \cite{gtrans}, which
can then be thought of as occupying the N lowest-lying
single-particle states labelled by the momenta $\{{\bf k}_\alpha\}$.
In Jain's original picture \cite{Jain89},
the determinant is just the wavefunction describing
$N$ filled LL's with $N\rightarrow \infty$, which is the limit $B=0$.

Read has suggested that the stability of the CF sea follows
from the low energy associated with the
Laughlin (bosonic) state, $\psi_2^{(0)}$, which places all the
zeros in the many-body wavefunction at the positions of the
particles \cite{ReadISI}.
The projection operator and the factors
$e^{ik_{\alpha} r_j}$
move one of the zeros in the wavefunction $\psi_2^{(0)}$,
considered as a function of the coordinate $z_j$,
away from the positions of the other
particles by an amount proportional to $k_\alpha$.
Filling the lowest momentum states therefore keeps
the zeros as close as possible to the particle positions
in an antisymmetric wavefunction. It is therefore natural
to associate the energy difference between the true ground state
energy for fermions and the energy $\psi_2^{(0)}$ with a
degeneracy temperature for the CF's.

To measure the stability of CF against
the `unbinding' of their flux tubes (zeros in the wavefunction),
we compare the energy of
$\psi_2^{(0)}$ to that of a system at the same density
but with pair correlation function corresponding to $\psi_1^{(0)}$.
A state with $N$ particles in the lowest LL
at the same filling fraction as $\psi_{CF}^{(0)}$
can be written as a product
$\psi_1^{(0)} P_{N-1}$, where $P_{N-1}$ is a symmetric polynomial
of order $(N-1)$.
In calculating the reference energy we write the polynomial
$P_{N-1} = \prod_{i,k}(z_i-\eta_k)$, and assume that the positions
of the $(N-1)$ zeros, $\eta_k$, are randomly distributed over the area
of the system. Using Laughlin's plasma analogy
\cite{Laughlin83},
we see that the zeros at $\eta_k$ act as point charges for
the electrons, so that averaging over their
positions just renormalizes the background charge in the plasma and
leads to the pair correlation function associated with $\psi_1^{(0)}$.
This reference energy is also the mean interaction energy at temperatures
larger than the condensation energy for the CF state but still small
with respect to the Zeeman energy and the LL splitting.

The state $\psi_{CF}^{(0)}$ is easily generalized to filling fractions
$\nu=1/2m$ by multiplying $\psi_{CF}^{(0)}$ by $\psi_{2m-2}^{(0)}$.
This state then has the maximum number, ($2m-1$), of zeros at the particle
positions and the remaining zero bound as closely as possible to
the particle positions consistent with fermion statistics.
Here, the CF state will only be stabilised, if it pays
energetically for the system to `bind' this remaining zero.
We therefore test the
energy of the CF `vacuum' state $\psi_{2m}^{(0)}$ against the
energy of a system at the same density
with pair correlation function associated with  $\psi_{2m-1}^{(0)}$.
Generalizing to higher LL's
(without accounting for LL-mixing) is also straightforward
\cite{MacDGirv86}.
We apply the product operator
$\alpha^+  =  \prod_i (a_i^+)$, where  $a_i^+$
is the LL raising operator for the $i$'th particle,
to the corresponding wavefunction
in the lowest LL:
\begin{equation}
\psi_{CF,m}^{(n)}   =  \left(\alpha^+ \right)^n \psi_{CF,m}^{(0)}
\label{eq:psiCFn}
\end{equation}
For a filling fraction of the $n$'th Landau level $\nu_n = 1/2m$, the
stability of the CF state should be determined by whether
the state $\psi_{2m}^{(n)}$ offers a significant gain in energy with
respect to the state at the same density with pair
correlation function from $\psi_{2m-1}^{(n)}$.
(The operation described by $\alpha^+$ is in fact simple to implement,
as the operators, $a_i^+$, leave all the intra-LL quantum numbers
unaffected.
In particular, there is a simple mapping associated with
$\alpha^+$ for the pair correlation function for electrons in the
$n$'th LL on the surface of a sphere, $g^{(n)}(\theta)$.
Here $\theta$ measures the spherical angle between two particles.
If we write
$g^{(n)}(\theta) = \sum_{\lambda = 0}^{2S} g_\lambda^{(n)}
Y_{\lambda 0}(\theta, \phi)$, where $2S$ is the flux through the
sphere and $Y_{\lambda m}$ are spherical harmonics, then
$g_\lambda^{(1)} = g^{(0)}_\lambda (1- \lambda(\lambda+1)/2S)$
\cite{Morf5/2}.)

We have calculated the energy of electrons on
a sphere described by the wavefunctions $\psi_2^{(n)}$ and $\psi_4^{(n)}$ for
both $n=0$ and $n=1$ using Monte Carlo simulations
with up to 24 particles, but ignoring LL-mixing effects.
We compare these with the energy of systems
described by $\psi_1^{(n)}$ and $\psi_3^{(n)}$ at the same density in
Table 1.
Both $\psi_2^{(0)}$ and $\psi_4^{(0)}$ describe systems with
significant energy gain with respect to the reference states
$\psi_1^{(0)}$ and $\psi_3^{(0)}$. The
gain is less for $\psi_4^{(0)}$ than for $\psi_2^{(0)}$ reflecting
the slow decay with angular momentum of the pseudopotentials characterizing
the Coulomb interaction \cite{Hald87}.
Taking the dielectric constant $\epsilon = 13$ for GaAs, the
gain for $\psi_2^{(0)}$ is equivalent to a temperature
4.6K for the density $\rho \sim 0.6 \times 10^{15}$m$^{-2}$ of the
samples used in \cite{Willett93}. This would be
consistent with the observation of CF-associated
conductivity anomalies up to temperatures $>4$K in those samples.

Table 1 shows that the picture in the $n=1$
LL is very different.
The energy of a system of particles described
by $\psi_2^{(1)}$ is in fact higher for the Coulomb interaction
than a system described by $\psi_1^{(1)}$ at the same density. On the
other hand $\psi_4^{(1)}$ has a significantly lower energy than
the reference state $\psi_3^{(1)}$.
These results suggest that
the CF state is stabilized at $\nu_1=1/4$ but should not
form for filling fraction $\nu_1=1/2$.
This would be in line with experiments \cite{Gammel88}  which
show evidence for the hierarchy of incompressible polarized states
at odd denominator
filling fractions $1/5 < \nu_1 < 1/3$, but not at
$1/3 < \nu_1 < 2/3$. (These incompressible states correspond
to filled Landau levels for CF's \cite{HLR,ambig}.) The high energy
associated with $\psi_2^{(1)}$ would also make unlikely the
formation of the $n=1$ counterparts of the
spin-singlet incompressible  fluids, like the
so-called `332' state at $\nu_0=2/5$. The stability of
these states is due basically to the low Coulomb energy associated
with the factor $\psi_2^{(0)}$ \cite{Halperin332}.

We can test the claim that a CF liquid state
will form at $\nu_1=1/4$ but not at $\nu_1=1/2$ by
considering the ground state angular momentum.
Rezayi and Read (RR) \cite{RR94} have shown that,
if the CF state is formed for a system of $N$ particles on a sphere
pierced by $2S = 2(N-1)$ flux units on a sphere,
then the angular momentum of the ground state should be the same as
that expected on the basis of Hund's second rule for
electrons in zero magnetic flux \cite{Jain94}.
We have calculated the ground state energy and angular momentum
for systems with up to nine particles at $\nu_n = 1/4$, ($2S = 4(N-n-1)$),
and up to twelve particles at $\nu_n = 1/2$, ($2S = 2(N-n-1)$), for
$n=0$ and $n=1$ (again ignoring LL-mixing).
At $\nu_0 = 1/2$, $\nu_0 = 1/4$
and  $\nu_1 = 1/4$, the angular momentum of the ground state
of the interacting system of N particles follows the RR prediction, L:
 \begin{quasitable}
 \begin{tabular}{cccccccccc} \hline
 N   & 4 & 5 & 6 & 7 & 8 & 9 & 10 & 11 & 12 \\ \hline
$L$  & 0 & 2 & 3 & 3 & 2 & 0 & 3 & 5 & 6 \\
$L'$ & 2 & 0 & 1 & 1 & 2 & 0 & 1 & 3 & 0 \\
\end{tabular}
\end{quasitable}
However, at $\nu_1 = 1/2$
there is no correlation between the ground state
angular momentum $L'$ and the value predicted by RR.
Taken together with the
high energy associated with $\psi_2^{(1)}$, this rules out the
CF state at $\nu_1 = 1/2$, when LL-mixing effects
are weak.

We have also checked the validity of the mapping (\ref{eq:psiCFn})
between CF ground states in different LL's. We evaluate the
overlap, $<\psi^{(1)}|\alpha^+\psi^{(0)}>$, where the
$\psi^{(n)}$ are the exact ground states in the $n$'th
LL for the Coulomb interaction with $\nu_1 = \nu_0$.
For the CF trial states (\ref{eq:psiCF}), (\ref{eq:psiCFn}) this
overlap is one while, for the exact ground states for a Coulomb
interaction, one would expect it to be large.
We find that, for $\nu_1=1/4$,
the overlap is larger than 0.975 for $4\leq N \leq 9$ particles,
while for $\nu_1 = 1/2$ it is either zero, when $L' \neq L$,
or very small.

The origin of the high energy of a system described by
$\psi_2^{(1)}$ can be seen in the small $r$ limit of the
pair correlation function, $g(r)$, \cite{Morf5/2}.
In Figure 1(a), we show $g(r)$
for $\psi_2^{(1)}$ and $\psi_1^{(1)}$. The Figure shows that,
whereas $g(r) \rightarrow 0$ as $r \rightarrow 0$ for $\psi_1^{(1)}$
(as it must on account of the Pauli principle),
$g(0)$ is non-zero and large for $\psi_2^{(1)}$. At
the same time we see in Figure 1(b) that, for
the state $\psi_4^{(1)}$, the probability
of finding two particles close together is reduced with respect to
that for the state $\psi_3^{(1)}$.

The pair correlation function for a many-body wavefunction
is determined by the amplitudes it ascribes to states of
relative angular momentum
foro pairs of particles. In the wavefunctions, $\psi_m^{(n)}$,
all configurations
in which pairs of particles have relative angular momentum
less than $m$ are excluded.
The reason why, in higher LL's, this does not lead to a small
probability that particles come close together,
is implicit in work on incompressible states
\cite{MacDGirv86,dARe88} but can be adapted to
our analysis of the CF states fairly easily.

The wavefunction for any pair of electrons can be written
in terms of their relative and centre of mass (CoM) coordinates,
$z$ and $Z$.
If both electrons are in the lowest LL then so is
their relative and CoM motion, and (in a
rotationally invariant system) the pair wavefunctions would
be of  the form
\begin{equation}
\psi^{(0)}_{jJ}(z_1,z_2) = \phi_j^{(0)}(z)  \Phi_J^{(0)}(Z) ,
\label{eq:relcom0}
\end{equation}
where $\phi_j^{(0)}$ and $\Phi_J^{(0)}$ describe
motion in the $n=0$ LL with relative and CoM angular momenta, $j$ and
$J$.
When the two electrons are both in the $n$'th LL,
the corresponding pair wavefunction is found by
acting with $(a_1^+ a_2^+)^n$
on the wavefunction $\psi_{jJ}^{(0)}$. In terms of
the raising operators for CoM and relative motion, $A^+$ and $a^+$,
we find
\begin{equation}
\psi^{(n)}_{jJ}(z_1,z_2) = \left(A^{+^2} - a^{+^2}\right)^n
                           \phi_j^{(0)}(z) \Phi_J^{(0)}(Z) ,
\label{eq:relcom1}
\end{equation}
so that the wavefunction for relative motion can be in any
of the 0, 2, 4, $\ldots 2n$'th LL's.
For all $j<2n$ with $j$ even, at least one of these $\phi_j^{(2l)}(z)$
is non-zero when $|z| \rightarrow 0$.
Thus for even $m$, excluding relative angular momentum less than $m$
in a many-body wavefunction (as in $\psi_m^{(n)}$), does
not suppress configurations in which pairs of
particles come close together unless  $m>2n$.

By matching the variation with system size of the ground state energy
to that expected for a CF liquid,
we can estimate the CF effective masses \cite{Morfetc}.
The ground state energy per particle of $N$ non-interacting
fermions on a sphere
is $(2\hbar^2/ma^2 N^2) \sum_i^N l_i(l_i+1)$.
Here $a$
is the ion disc radius for the particles of mass
$m$, and
the $l_i$ are the angular momenta of the $N$ lowest energy
single-particle states. This energy can be written
$(\hbar^2/ma^2)(1-\delta(N))$,
where $\delta(N) = 1/N$,
for filled shells, and $0 \leq \delta(N) < 1/N$ otherwise.
The variation with system size in a sequence of $L=0$ ground states
at fixed filling fraction is also known to be proportional to $1/N$
\cite{dAMorf89}, where
the $1/N$ corrections for particles on a sphere relate to its curvature
and depend on the precise definition of the interparticle separation
\cite{MorfHalp87}. We would like to eliminate such background corrections
to obtain a best estimate of the systematic variation of
the ground state energy with angular momentum.
We therefore subtract a `filled
shell' equivalent
energy, $E_{fs}(N) = e_0 - e_1/N$,
from the ground state energy for each $N$, and compare the result with
the corresponding result, $(\hbar^2/ma^2)(\delta(N)-1/N)$,
expected for free fermions.
We use the $N=4$ and $N=9$ particle ground state energies to determine
the coefficients $e_0$ and $e_1$.
The results at $\nu_0 = 1/2$ and $\nu_1=1/4$ are shown in Figure 2.

The close agreement between the energies at $\nu_0 = 1/2$ and
those of the non-interacting fermions apparent from Figure 2 allows us
to estimate effective masses of the CFs. We find that,
in the relation \cite{HLR}:
\begin{equation}
\frac{\hbar^2}{m^*} = \frac{C}{(4\pi \rho^{(n)})^{1/2}}
\frac{e^2}{\epsilon},
\label{eq:mstar}
\end{equation}
where $\rho^{(n)}$ is the number density of electrons in the $n$'th LL and
$\epsilon$ is the dielectric constant,
the constant $C = 0.20 \pm 0.02$.
This differs from the
result, $C \approx 0.31$, quoted in \cite{HLR}.
However, this may relate to the
enhancement expected for $m^*$ as $\nu \rightarrow 1/2$.
(Our results are for systems at $\nu=1/2$,
whereas the estimate of \cite{HLR} was based on the scaling of gap
energies with filling fraction for incompressible states
away from $\nu = 1/2$.)
Similar analyses give
$C \approx 0.1 \pm 0.02$ at $\nu_0=1/4$ and $C \approx 0.18 \pm 0.02$ at
$\nu_1 = 1/4$, although these figures
ignore the low ground state energies for six and seven particles
which we find for both $n=0$ and $n=1$.

We thank the ISI Foundation in Torino for their
hospitality and the participants
in the ISI workshop on the Quantum Hall Effect, June 1994, for
many interesting seminars.
We also thank B.I. Halperin and E. Rezayi for useful
discussions.


  \newpage

\figure{
The pair correlation function, $g(r)$, for systems
of nine particles.
The upper panel shows $g(r)$ for systems
described by $\psi_1^{(1)}$,  $\psi_2^{(1)}$
(see \ref{eq:psiCFn}) and by the exact (Coulomb)
ground state wavefunction at $\nu_1 = 1/2$.
The lower panel shows $g(r)$ for $\psi_3^{(1)}$, $\psi_4^{(1)}$
and for the exact ground state wavefunction at $\nu_1 = 1/4$.
The large value of $g(r)$ as
$r \rightarrow 0$ for $\psi_2^{(1)}$ leads to a high energy per
particle (Table \ref{Ebind}), and makes the formation of
CF's energetically unfavourable. However,
$\psi_4^{(1)}$ has significantly more weight at small $r$ than the
reference state $\psi_3^{(1)}$. This leads to a large energy gain
for the binding of the additional zero in the wavefunction and
implies a well-stabilised CF-liquid state at $\nu_1 = 1/4$.
}

\figure{
The ground state energy minus the respective `filled shell' equivalent
energies for non-interacting free fermions and for
interacting electrons at $\nu_0=1/2$ and at $\nu_1=1/4$
on the surface of a sphere.
The constant $C$ in (\ref{eq:mstar}) is taken as 0.2 at $\nu_0=1/2$
and 0.18 at $\nu_1=1/4$.
(In units of $e^2/\epsilon a^{(n)}$,
the filled shell equivalent energies for electrons are
$E_{fs}(N)
 = -0.9321 - 0.0830/N$ at $\nu_0=1/2$,
$E_{fs}(N)
 = -1.0205 - 0.0028/N$ at $\nu_0=1/4$ and
$E_{fs}(N)
 =  -0.8825 + 0.0047/N$ at $\nu_1=1/4$.)
}

\newpage

 \begin{table}

 \caption[]{The energy per particle of a system of particles described by
$\psi_m^{(n)}$. The energy difference,
$E_{cf}^{(n)}(\frac{1}{2m}) = E(\psi_{2m}^{(n)}) - E(\psi_{2m-1}^{(n)})$,
is a measure of the stability of a CF state. The positive figure
for $\nu_1 = 1/2$ suggests that the CF state
will not form at this filling fraction. All energies are in units
of $e^2/\epsilon a^{(n)}$ where $a^{(n)}=1/(\pi\rho^{(n)})^{1/2}$
is the ion disc radius for the electrons in
the $n$'th LL with density $\rho^{(n)}$.

}

 \label{Ebind}

 \begin{tabular}{c|ccc|ccc} \hline

$n$ & $\psi_1^{(n)}$ & $\psi_2^{(n)}$ & $E_{cf}^{(n)}(\frac{1}{2}) $ &
      $\psi_3^{(n)}$ & $\psi_4^{(n)}$ & $E_{cf}^{(n)}(\frac{1}{4}) $
\\ \hline
0 & -0.886 & -0.970 & -0.084  & -1.004 & -1.022 & -0.018 \\
1  & -0.664 & -0.502 & +0.162 & -0.799 & -0.889 &  -0.090 \\
 \hline
 \end{tabular}

 \end{table}



\begin{references}
\bibitem{ambig} There is an ambiguity associated with
incompressible states at $\nu_n = 1/(2m+1)$ in the CF picture as
these can be related to filled LL's of CF's at $\nu_n = 1/2m$ and
$\nu_n = 1/(2m+2)$. Clearly the $\nu_1 = 1/3$ incompressible state
\cite{Gammel88} can only be associated with the CF liquid at $\nu_1=1/4$ if
the CF state  does  not form at $\nu_1=1/2$.
\end{references}
\end{document}